\documentclass[
prd,
twocolumn,
aps,
preprintnumbers,
nofootinbib,
showkeys]
{revtex4}

\usepackage{latexsym}
\usepackage{amsfonts}
\usepackage{amssymb}
\usepackage{amsmath}
\usepackage{graphicx}
\usepackage{epic}
\usepackage{eepic}
\usepackage{epsfig}
\usepackage{float}
\usepackage{color}
\usepackage{multirow}
\usepackage{array}
\usepackage{url}  
\usepackage[caption=false]{subfig}
\usepackage{hyperref}
\hypersetup{colorlinks = true, linkcolor = blue, urlcolor  = blue, citecolor = blue, anchorcolor = blue}
\usepackage{bm}

\usepackage{shorthand}

\begin{document}

\title{LHC probes of TeV-scale scalars in $\mathrm{SO}(10)$ grand unification}

\author{Ufuk Aydemir} 
\email{ufuk.aydemir@physics.uu.se}

\author{Tanumoy Mandal} 
\email{tanumoy.mandal@physics.uu.se}

\address{Department of Physics and Astronomy,
Uppsala University, Box 516, SE-751 20 Uppsala, Sweden}


\begin{abstract}
We investigate the possibility of TeV-scale scalars as low energy remnants arising in the 
non-supersymmetric $\mathrm{SO}(10)$ grand unification framework where the field content is minimal. 
We consider a scenario where the $\mathrm{SO}(10)$ gauge symmetry is broken into the gauge symmetry 
of the Standard Model (SM) through multiple stages of symmetry breaking, and a colored and hypercharged scalar $\chi$ picks a TeV-scale mass in the process. The last stage of the symmetry breaking occurs at the TeV scale where the left-right symmetry, i.e. $\mathrm{SU}(2)_L\otimes \mathrm{SU}(2)_R\otimes \mathrm{U}(1)_{B-L}\otimes \mathrm{SU}(3)_C$, is broken into that of the SM by a singlet scalar field 
$\mathcal{S}$ of mass $M_{\mathcal{S}}\sim 1$ TeV, which is a component of an $\mathrm{SU}(2)_R$-triplet scalar field, acquiring a TeV-scale vacuum expectation value. For the LHC phenomenology, we consider a scenario where $\mathcal{S}$ is produced via gluon-gluon fusion through loop interactions with $\chi$ and also decays to a pair of SM gauge bosons through $\chi$ in the loop. We find that the parameter space is heavily constrained from the latest LHC data. We use a multivariate analysis to estimate the LHC discovery reach of $\mathcal{S}$ into the diphoton channel. 
\end{abstract}
 

\maketitle

\section{Introduction\label{sec:intro}}

After the discovery of the Higgs boson at the Large Hadron Collider (LHC)~\cite{Chatrchyan:2012xdj,Aad:2012tfa}, the last piece of the triumphant achievement of the high energy physics community - the Standard Model (SM); the great expectations for the observation of some sort of new physics at the LHC, emanated from the paradigms based on the familiar intuitions, some of which have so far lead the community to success, have turned out to be great disappointments as the LHC searches to date have returned empty-handed. Although there have been a couple of noticeable excesses, such as the diphoton~\cite{Aaboud:2016tru,Khachatryan:2016hje} (see Ref.~\cite{Strumia:2016wys} for a review and the full list of references) and diboson \cite{Aad:2015owa,Aad:2014xka,Aad:2015ufa} anomalies, which caused excitement among the community, these signals have turned out to be statistical fluctuations as more data accumulates in. 

While the LHC is still up and running, and looking for any hint of trace pointing to physics beyond the SM (BSM), the community has been in an ambitious effort for projecting out the LHC implications of variety of new physics models for a possible future discovery. Among the various search channels, the diphoton resonance search is one of the most important
programs at the LHC since this channel provides a comparatively cleaner background. 
One of the key predictions of many BSM theories is the existence of diphoton resonances around the TeV-scale arising from the decay of 
TeV-scale scalars present in those models.

One of the most appealing scenarios for a more fundamental picture is the Grand Unified Theory (GUT) framework, in which the $\mathrm{SO}(10)$ GUT is particularly interesting~\cite{Chang:1983fu,Chang:1984uy,Chang:1984qr,Parida:1989an,Deshpande:1992au,Bajc:2005zf,Bertolini:2009qj,Babu:2012vc,Awasthi:2013ff,Nayak:2013dza,Parida:2014dba,Brennan:2015psa,Aydemir:2015nfa,Bandyopadhyay:2015fka,Aydemir:2015oob,Parida:2016hln} (see Refs.~\cite{Deshpande:1992eu,Fukuyama:2004xs,Fukuyama:2004ps,Majee:2007uv,Parida:2008pu,Dev:2009aw,Parida:2010wq} for analyses of the supersymmetric $\mathrm{SO}(10)$ GUT).
Breaking the $\mathrm{SO}(10)$ gauge symmetry into that of the SM can be realized in a single step as well as in multiple steps by various symmetry breaking sequences. The relevant option we consider in this paper is the latter, while one possible intermediate phase, which we assume to be in the TeV-scale, is the left-right model whose gauge symmetry is based on $\mathrm{SU}(2)_L\otimes \mathrm{SU}(2)_R\otimes \mathrm{U}(1)_{B-L}\otimes \mathrm{SU}(3)_C$ 
($G_{2213}$)~\cite{Pati:1974yy,Mohapatra:1974gc,Senjanovic:1975rk,Mohapatra:1979ia,Mohapatra:1980yp,Duka:1999uc,Aydemir:2013zua,Aydemir:2014ama}, which is different than the left-right symmetric version since in this case $\mathrm{SU}(2)_L$ and $\mathrm{SU}(2)_R$ gauge couplings are different i.e. $g_L\neq g_R$. Adopting the minimalistic approach and therefore, keeping the initial field content 
(the $\mathrm{SO}(10)$ multiplets) minimal, and tempted by the least possible fine-tuning intuition, it seems not possible to obtain a plausible scenario where the left-right model lies in the TeV-scale~\cite{Aydemir:2015oob}. For instance, if the Higgs content is determined based on the \textit{extended survival hypothesis} (ESH)~\cite{delAguila:1980qag}, the model does not allow symmetry breaking scale of the left-right model to be in the TeV-scale. Recall that the ESH states that at every step of a symmetry breaking sequence, the only scalars which survive below the corresponding symmetry breaking scale are the ones which acquire vacuum expectation values (VEVs) at the subsequent levels of the
symmetry breaking. However, by slightly relaxing the ESH conjecture by allowing one or more colored scalars to become light (at the TeV scale), it is possible to have a TeV scale left-right model in the 
$\mathrm{SO}(10)$ framework~\cite{Aydemir:2015oob}. 


In this paper, we investigate the phenomenology of TeV-scale scalars as low energy remnants
of the non-supersymmetric $\mathrm{SO}(10)$ GUT. The part of the model that lies in the TeV-scale, as mentioned above, is the left-right model, augmented by a color-triplet scalar $\Delta_R (1,3,2/3,3)$, whose one component $\chi$, we assume for our demonstration, has a mass of $\sim1$ TeV, while its other components are heavier in the TeV range. In particular, we explore the phenomenology of a SM-singlet scalar 
$\mathcal{S}$ of mass around 1 TeV which is assumed to be the excitation of the neutral component of an $\mathrm{SU}(2)_R$ triplet $\Delta_{R_1}(1,3,2,1)$, denoted as $\Delta_{R_1}^{0}$. The field $\Delta_{R_1}^{0}$ breaks the symmetry of the left-right model into that of the SM by acquiring a VEV presumably at the 
TeV-scale in our set-up. The scalar $\chi$ is responsible for the production and decay
of $\mathcal{S}$ through loop interactions.  

In our model, we assume two intermediate energy scales between the electroweak scale $M_Z$ and the unification scale $M_U$. At the scale $M_U$, the $\mathrm{SO}(10)$ is broken into the Pati-Salam group, $\mathrm{SU}(2)_L\otimes \mathrm{SU}(2)_R\otimes \mathrm{SU}(4)_C$ ($G_{224}$). The Pati-Salam group is broken into the group of the left-right model at the first intermediate energy scale $M_C$, which is followed by the breaking of the left-right model into the SM at the energy scale $M_R$. In our scenario, $M_R$ is assumed to be in the TeV scale, while the values of $M_U$ and $M_C$ come out as predictions of the model. Note that the $D$-parity invariance \cite{Chang:1984uy,Chang:1984qr,Maiezza:2010ic}, which is a $Z_2$ symmetry that maintains the complete equivalence of the left and the right sectors, is broken together with the $\mathrm{SO}(10)$ in the first stage of the symmetry breaking. Therefore, the gauge couplings associated with the $\mathrm{SU}(2)_L$ and 
$\mathrm{SU}(2)_R$ gauge groups, $g_L$ and $g_R$, evolve under the influence of different particle contents, hence $g_R\neq g_L$, below the scale $M_U$. Remember that the $D$-parity is slightly different from the usual Lorentz parity in that the latter does not transform scalars, while the $D$-parity transforms them non-trivially. Note also that we remain in the minimal picture in terms of the total field content; the model does not have any extra matter field or any scalar 
$\mathrm{SO}(10)$ multiplet other than the ones required to begin with. Thus, the advantage of having a TeV-scale colored scalar is two-fold: it is responsible for the
production and decays of $\mathcal{S}$ and it can successfully be embedded in the minimal non-supersymmetric $\mathrm{SO}(10)$ GUT scheme while maintaining the field content minimal. 

In this paper, we identify the region of parameter space of our model constrained from the latest LHC data. By using a multivariate analysis (MVA), we compute the higher-luminosity LHC discovery reach of 
$\mathcal{S}$ into the diphoton channel where, as we will discuss later, the most stringent bounds come from.

The paper is organized as follows. In Section~\ref{sec:mod}, we review the left-right model in the 
$\mathrm{SO}(10)$ grand unification framework. We discuss how the two scalars, 
$\mc{S}$ and $\chi$ of our interest, arise in our set-up. In Section~\ref{sec:unification}, we discuss the unification of the couplings, derive the values of the intermediate symmetry breaking scales, and present the resulting predictions of the model. In Section~\ref{sec:pheno}, we present the phenomenology
of $\mc{S}$ and $\chi$ including the exclusion limits from the LHC data and future discovery prospects. We summarize our conclusions in Section~\ref{summary}.  


\section{The Model}
\label{sec:mod}
We consider a left-right model, whose gauge group is 
$\mathrm{SU}(2)_L\otimes \mathrm{SU}(2)_R\otimes \mathrm{U}(1)_{B-L}\otimes \mathrm{SU}(3)_C$, which is assumed to be broken into the SM at the TeV scale. 
The breaking is realized by the neutral component ($\Delta_{R_1}^{0}$ which we denote as $\mc{S}$) of the 
$\mathrm{SU}(2)_R$ triplet $\Delta_{R_1}(1,3,2,1)$, which is commonly preferred in the 
literature. Here, instead of the $\mathrm{SU}(2)$ triplets, the $\mathrm{SU}(2)$ doublet 
$(1,2,1,1)$, which originates from the $\mathrm{SO}(10)$ multiplet $\textbf{16}$, can also be used. The advantage of the triplet representation is that it can provide a Majorana mass term for the right-handed neutrino and hence, the seesaw mechanism~\cite{Minkowski:1977sc,GellMann:1980vs,Yanagida:1979as,Schechter:1980gr,Schechter:1981cv} for small neutrino masses.
 
In this work, we explore the phenomenology of the SM-singlet $\mc{S}$ 
which we assume to be produced and decayed through the loop interaction with a color-triplet hypercharged scalar $\Delta_{R_3}^{4/3}\left(1,8/3,3\right)$ denoted as $\chi$. The $\chi$ originates from the decomposition of 
the $\Delta_R (1,3,2/3,3)$ component of the $G_{224}$ multiplet $\Delta_R (1,3,10)$ into the SM group as follows
\begin{align}
\Delta_{R_3}\left(1,3,\dfrac{2}{3},3\right)&=\Delta_{R_3}^{\frac{4}{3}}\left(1,\dfrac{8}{3},3\right)\oplus\Delta_{R_3}^{\frac{1}{3}}\left(1,\dfrac{2}{3},3\right) \nn\\
&\oplus\Delta_{R_3}^{-\frac{2}{3}}\left(1,\dfrac{-4}{3},3\right)\ .
\label{eq:triplet}
\end{align}
For our purpose, we take the mass of $\chi$ around 1 TeV, while the other components have heavier masses, $\sim 2-5$ TeV, and hence, their contribution to the production and the decay 
of $\mc{S}$ are relatively suppressed. 

The SM electroweak symmetry breaking (EWSB) in the left-right model, in general, is achieved by the neutral (diagonal) component of the bidoublet field $\phi(2,2,0,1)$ acquiring a VEV.
The fermion content of the model is the same as the SM.
There are seven gauge bosons in the model, $W_L^i$, $W_R^i$ (with $i=1,2,3$) and $W_{BL}$, with the gauge couplings $g_L$, $g_R$. and $g_{BL}$, associated with the $\mathrm{SU}(2)_L$,  $\mathrm{SU}(2)_R$, and $\mathrm{U}(1)_{B-L}$ gauge symmetries, respectively.
Using the notation of Refs.~\cite{Aydemir:2015nfa,Aydemir:2015oob}, the symmetry breaking pattern of our model is given by
\begin{align}
\label{chain}
\mathrm{SO}(10)\underset{\left<\mathbf{210}\right>}{\xrightarrow{M_U}} G_{224} 
\underset{\left<\mathbf{210}\right>}{\xrightarrow{M_C}}G_{2213}\underset{\left<\mathbf{126}\right>}{\xrightarrow{M_R}}G_{213}\underset{\left<\mathbf{10}\right>}{\xrightarrow{M_Z}}G_{13}\;,
\end{align}
where we assume $M_R=5$ TeV in our analysis.

In choosing the $\mathrm{SO}(10)$ multiplets for breaking the symmetries (by acquiring appropriate VEVs), we follow the common tradition in the literature as follows. The first stage of the symmetry breaking, where $\mathrm{SO}(10)$ is broken into the Pati-Salam group $G_{224}$, is realized by the singlet $(1,1,1)_{210}$ of $\textbf{210}$. Note that $(1,1,1)_{210}$ is odd under the $D$-parity~\cite{Chang:1984uy,Chang:1984qr}, and hence, it is broken at this stage as well. Therefore, below the scale $M_U$, we have $g_L\neq g_R$, since they evolve under the influence of different particle contents below this energy scale according to the ESH and the minimal fine tuning principle. The second stage, where the Pati-Salam group is broken into the left-right group $G_{2213}$, can be accomplished by $(1,1,15)_{210}\equiv \Sigma(1,1,15)$ acquiring a VEV. The breaking of  $G_{2213}$ down to the SM gauge group $G_{213}$ is achieved by the $G_{2213}$ multiplet $(1,3,2,1)_{126}\equiv \Delta_{R_1}(1,3,2,1)$ which belongs to the Pati-Salam multiplet $(1,3,10)_{126}\equiv \Delta_R (1,3,10)$ which is a member of the 
$\mathrm{SO}(10)$ multiplet $\textbf{126}$. In our model, $\Delta_{R_1}(1,3,2,1)$ acquires a VEV at around $5$ TeV which also set the value of the symmetry breaking scale $M_R$. Note that $\Delta_{R_1}$ is the regular $\mathrm{SU}(2)_R$ triplet usually used in the literature in order to break the $G_{2213}$ symmetry.

\section{Unification of the couplings}
\label{sec:unification}

In this section, we discuss how the unification of the couplings are achieved and derive the values of the symmetry breaking scales. We have only two intermediate scales in our model in between the unification scale $M_U$ and the EWSB scale $M_Z$, which are $M_C$ and $M_R$, where the value of $M_R$ is chosen to be 5 TeV.  

The TeV-scale left-right model with light colored scalars in the minimal non-supersymmetric $\mathrm{SO}(10)$ GUT scheme has recently been discussed in~\cite{Aydemir:2015oob}. Here, the situation has a slight difference in that one of the components in the decomposition of the left-right multiplet $\Delta_{R_3}$ (shown in Eq.~\eqref{eq:triplet}) into the SM gauge group, which is $\Delta_{R_3}^{4/3}$ whose mass is $\sim 1$ TeV. Therefore, the renormalization group (RG) running of the gauge couplings at this energy scale is slightly different. The other particle $\mc{S}$ which, we assume, has a mass also around $\sim 1$ TeV, does naturally not contribute to the running since it is a SM-singlet.

\subsection{Basics} 

We label the energy intervals in between symmetry breaking scales
starting from $[M_Z,M_R]$ up to $[M_C,M_U]$ with Roman numerals as:
\begin{align}
\mathrm{I} : \underbrace{[M_Z,M_R]}_{G_{213}}\ ;~~
\mathrm{II} : \underbrace{[M_R,M_C]}_{G_{2213}}\ ;~~
\mathrm{III} : \underbrace{[M_C,M_U]}_{G_{224}}\;.
\label{IntervalNumber}
\end{align}
The boundary/matching conditions we impose on the couplings at the symmetry breaking scales are:
\begin{align}
M_U &: g_L(M_U) \;=\; g_R(M_U) \;=\; g_4(M_U) \;,\label{MUmathcing}\\
M_C &: \sqrt{\frac{2}{3}}\,g_{BL}(M_C) \;=\; g_3(M_C)=g_4(M_C) \;,\label{MCmatching}\\
M_R &: \frac{1}{g_1^2(M_R)} \;=\; \frac{1}{g_R^2(M_R)}+\frac{1}{g_{BL}^2(M_R)}\nn\\
&~~\textrm{and}~~g_2(M_R)\;=\;g_L(M_R)\;, \label{MRmatching} \\
M_Z &: \frac{1}{e^2(M_Z)} \;=\; \frac{1}{g_1^2(M_Z)}+\frac{1}{g_2^2(M_Z)}\;.
\label{MZmatching}
\end{align}
The low energy data which we will use as boundary conditions to the RG running are
\cite{Olive:2016xmw,ALEPH:2005ab}
\begin{align}
\alpha = 1/127.9;~~\alpha_s = 0.118;~~\sin^2\theta_W = 0.2312,
\label{SMboundary}
\end{align}
all are evaluated at $M_Z=91.2$ GeV, which gives
\begin{align}
g_1(M_Z)= 0.36,~~g_2(M_Z) = 0.65,~~g_3(M_Z) = 1.22\ .
\label{MZboundary}
\end{align}
Note that the coupling constants are all required to remain in the perturbative regime during the
evolution from $M_U$ down to $M_Z$.

\subsection{One-loop RG running}

For a given particle content; the gauge couplings, in an energy interval $\lt[M_A,M_B\rt]$, are evolved  according to the one-loop RG relation
\begin{align}
\frac{1}{g_{i}^{2}(M_A)} - \dfrac{1}{g_{i}^2(M_B)}
\;=\; \dfrac{a_i}{8 \pi^2}\ln\dfrac{M_B}{M_A}
\;,
\end{align}
where the RG coefficients $a_i$ are given by \cite{Jones:1981we,Lindner:1996tf} as
\begin{align}
\label{1loopgeneral}
a_{i} &= -\frac{11}{3}C_{2}(G_i)
 + \frac{2}{3}\sum_{R_f} T_i(R_f)\cdot d_1(R_f)\cdots d_n(R_f) \nn\\
& + \frac{\eta}{3}\sum_{R_s} T_i(R_s)\cdot d_1(R_s)\cdots d_n(R_s)\;.
\end{align}
Here, the two summations are over irreducible chiral representations of fermions $R_f$ and those of scalars $R_s$. The coefficient $\eta$ is either 1 or 1/2, depending on whether the representation is complex or real, respectively. The quadratic Casimir for the adjoint representation of the group $G_i$ is $C_2(G_i)$
and $T_i$ is the Dynkin index of each representation. For $\mathrm{U}(1)$ group, $C_2(G)=0$ and
\begin{equation}
\sum_{f,s}T = \sum_{f,s}\left(\dfrac{Y}{2}\right)^2\;,
\label{U1Dynkin}
\end{equation}
where $Y/2$ is the $\mathrm{U}(1)$ charge, the factor of $1/2$ coming from the traditional
normalizations of the hypercharge and $B-L$ charges.
The $a_i$'s differ depending on the particle content in each energy interval, which changes every time symmetry breaking occurs. We will distinguish the $a_i$'s in different intervals with the corresponding roman numeral superscript, cf. Eq.~\eqref{IntervalNumber}.

\subsection{Results}

The scalar content in the energy intervals are:
\begin{align}
\mathrm{III} &: \phi(2,2,1),~\Delta_R (1,3,10),~\Sigma(1,1,15)\;,\nonumber\\
\mathrm{II}  &: \phi(2,2,0,1),~\Delta_{R_1}(1,3,2,1),~\Delta_{R_3}\left(1,3,\dfrac{2}{3},3\right),~\vphantom{\bigg|}\nonumber\\
\mathrm{I}   &:\phi_2(2,1,1),~\mc{S}(1,1,1),~\Delta_{R_3}^{4/3}\left(1,\dfrac{8}{3},3\right)\;.
\end{align}
It is common in the literature that another scalar Pati-Salam multiplet, $\widetilde{\Sigma} (2,2,15)$, is included in interval III for a rich Yukawa phenomenology~\cite{Bajc:2005zf,Bertolini:2009qj}. In terms of the RG evolution, which is our main focus here, this extra multiplet would not alter the results noticeably, because its effect in the RG equations would appear as a contribution in the term $(-5 a_L+3 a_R +2 a_4)$ (see Eqs.~\eqref{A3}, which would be very small compared to the rest of the term. Therefore, for the sake of staying minimal, we do not include this multiplet in our set-up.

The values of the RG coefficients for this Higgs content are listed in Table~\ref{a1}.
The relations between symmetry breaking scales, which can be derived by using the one-loop running equations and the boundary/matching conditions, can be obtained as (for derivation see Refs.~\cite{Aydemir:2015nfa,Aydemir:2015oob})
\begin{align}
\label{A3}
2\pi\left[\dfrac{3}{\alpha} - \dfrac{8}{\alpha_s}\right] 
&= \left(3a_L+3a_R-6a_4\right)^\mathrm{III}\ln\dfrac{M_U}{M_C}\nn\\
&+ \left(3a_L + 3a_R + 3a_{BL} - 8a_3\right)^\mathrm{II}\ln\frac{M_C}{M_R}\nn\\
&+\left(3a_1 + 3a_2 - 8a_3\right)^\mathrm{I}\ln\dfrac{M_R}{M_Z}\ ,\\
2\pi\left[\dfrac{3-8s_w^2}{\alpha}\right]
&= \left(-5a_L+3a_R+2a_4\right)^\mathrm{III}\ln\dfrac{M_U}{M_C}\nn\\
&+ \left(-5a_L + 3a_R + 3a_{BL}\right)^\mathrm{II}\ln\dfrac{M_C}{M_R}\nn\\
&+ \left(3a_1 -5a_2\right)^\mathrm{I}\ln\dfrac{M_R}{M_Z}\ ,
\end{align}
where $s_w\equiv \sin\theta_W$.
Using these equations and the experimentally measured quantities in Eq.~\eqref{SMboundary}, and demanding $M_R=5$ TeV, we obtain the following values,
\begin{align}
\label{scales}
M_C= 10^{15.0}~\mbox{GeV}~~\mbox{and}~~M_U=10^{17.9}~~\mbox{GeV}\;.
\end{align}
The value for the scale $M_C$ is sufficiently high to ensure that the effects induced by the presence of scalar and vector-leptoquarks are suppressed adequately enough to remain consistent with the experimental constraints~\cite{Evans:2015cqq}. Besides, the unification scale $M_U$ is high enough to escape the bound on the proton decay induced by gauge boson exchanging operators. We should also note that we have light color-triplets in our model, and as well known they lead to scalar-induced dimension-6 operators that contribute to the proton decay amplitude. Although these contributions are typically suppressed by small Yukawa couplings, the color-triplets being as light as the TeV-scale can cause a potentially dangerous situation~\cite{Bertolini:2012im}. In such a case, a mechanism is required to adequately suppress these interactions, such as the ones proposed in Refs.~\cite{Dvali:1995hp,Rakshit:2003wj}.

\begin{figure}
\centering
\includegraphics[scale=0.6]{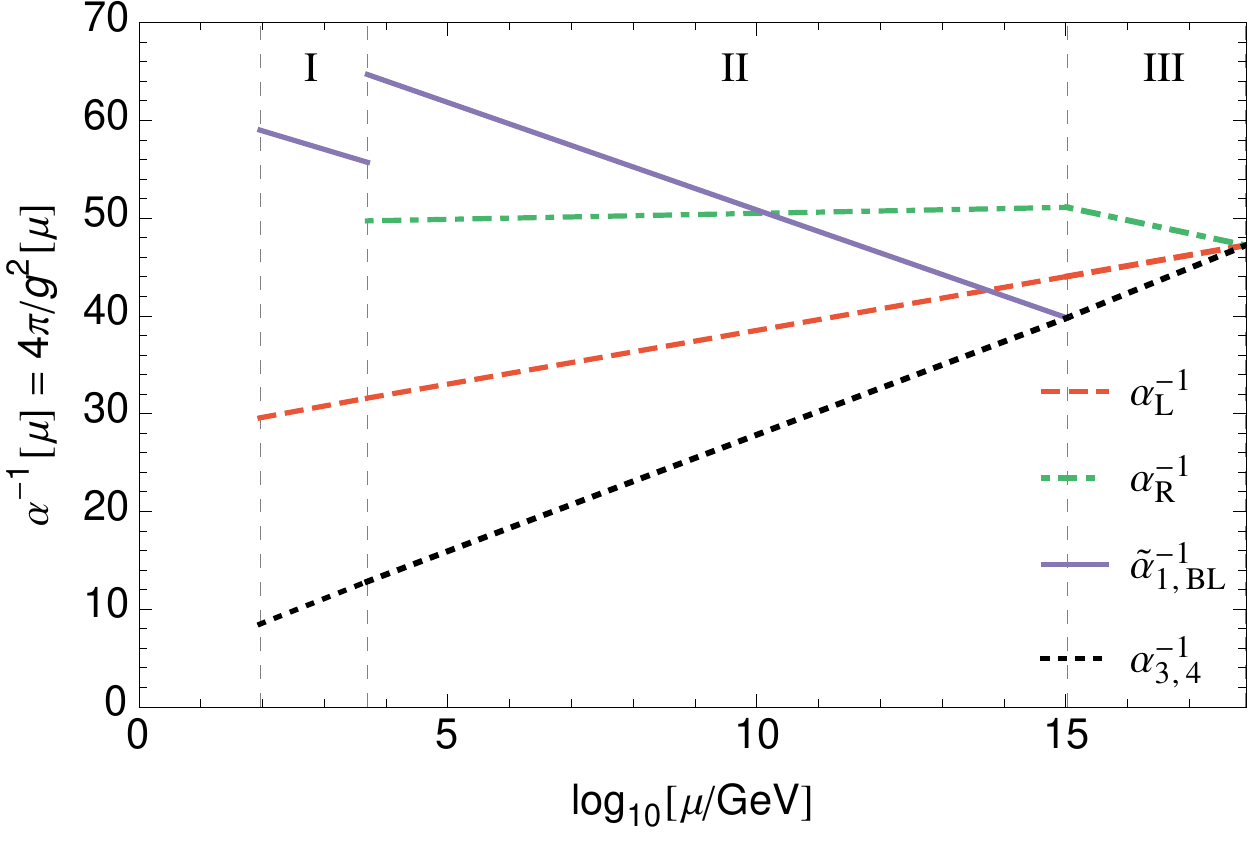}
\caption{Running of the gauge couplings for the model. The vertical dotted lines from left to right correspond to the
symmetry breaking scales $M_Z$, $M_R$, and $M_C$,  which also indicate the beginning of the energy intervals I, II, and III, respectively. For $\alpha_1^{-1}$ and $\alpha^{-1}_{BL}$, we plot the redefined quantities $\widetilde{\alpha}^{-1}_1\equiv (3/5)\alpha^{-1}_1$ and $\widetilde{\alpha}^{-1}_{BL}\equiv (3/2)\alpha^{-1}_{BL}$.  Note that the discontinuity on the $\widetilde{\alpha}^{-1}_{1,BL}$ plot at the energy scale $M_R$ occurs due to the boundary condition given in Eq.~(\ref{MRmatching}).} 
\label{fig:running}
\end{figure}

\begin{table*}
\centering
{\begin{tabular}{c|c|c}
\hline
$\vphantom{\Big|}$ Interval & Higgs content & RG coefficients
\\
\hline
$\vphantom{\Biggl|}$ III
& $\phi(2,2,1),\;\Delta_R (1,3,10),\;\Sigma(1,1,15)$
& $\left( a_{L},a_{R},a_{4}\right)^\mathrm{III}
 =\left(-3,\dfrac{11}{3},-7\right)$
\\
\hline
$\vphantom{\Biggl|}$  II
& $\phi(2,2,0,1),\;\Delta_{R_1}(1,3,2,1),\;\Delta_{R_3}\left(1,3,\dfrac{2}{3},3\right)$
& $\left(a_{L},a_{R},a_{BL},a_3\right)^\mathrm{II}=\left(-3,\dfrac{-1}{3},4,\dfrac{-13}{2}\right)$ $\vphantom{\Bigg|}$
\\
\hline
$\vphantom{\Biggl|}$   I
& $\phi_2(2,1,1),\; \mc{S}(1,1,1),\;\chi\left(1,\dfrac{8}{3},3\right)$
& $\left(a_{1},a_{2},a_{3}\right)^\mathrm{I}=\left(\dfrac{155}{18},\dfrac{-19}{6},\dfrac{-41}{6}\right)$
\\
\hline
\end{tabular}}
\caption{\label{a1} The Higgs content and the RG coefficients in the energy intervals for our model.}
\end{table*}
The value of the unified gauge coupling can be found via the following equation
\begin{align}
\label{A6}
\dfrac{2\pi}{\alpha_s}-\dfrac{2\pi}{\alpha_U}= 
a_4^\mathrm{III}\;\ln\dfrac{M_U}{M_C}
+ a_3^\mathrm{II}\;\ln\dfrac{M_C}{M_R}
+ a_3^\mathrm{I}\;\ln\dfrac{M_R}{M_Z}
\end{align}
as $\alpha^{-1}_U\simeq 47.2$.
The running of the couplings are given in Fig.~\ref{fig:running}.
Similarly, the gauge couplings at $M_R=5$ TeV are obtained as
\begin{align}
\label{coupling}
g_R\simeq 0.50;~~g_L\simeq 0.63;~~g_{BL}\simeq 0.55;~~g_3\simeq 0.99\ ,
\end{align}
which, together with the values of the symmetry breaking scales in Eq.~\eqref{scales}, are the main predictions of the model. Notice that the value of $g_R(5~\mbox{TeV})$ is different from the value of 
$g_2(5~\mbox{TeV})=g_L(5~\mbox{TeV})\simeq 0.63$, which is expected due to the fact that the $D$-parity invariance is broken together with the $\mathrm{SO}(10)$ symmetry, hence $g_R\neq g_L$  below the unification scale $M_U$, as mentioned previously.
The model also predicts the existence of TeV-scale gauge bosons $W_R$ and $Z_R$ whose masses at
$M_R$ are given as
\begin{align}
M_{W_R}\approx g_R v_R;~~
M_{Z_R}\approx \sqrt{2\lt(g_R^2 + g_{BL}^2\rt)}~v_R\;,
\end{align}
where we choose $v_R\equiv M_R=\left<\Delta_{R_1}^0\equiv \mc{S} \right>= 5$ TeV which, together with Eq.~(\ref{coupling}), yields
\begin{align}
\label{Wpr}
M_{W_R}(M_R)\approx 2.5~\mbox{TeV}~\mbox{and}~M_{Z_R}(M_R)\approx 5.3~\mbox{TeV}\;.
\end{align}
These are the specific predictions of our model. However, we note that $M_{W_R}$ and 
$M_{Z_R}$ change significantly with the choice of the symmetry breaking scale $M_R$. Therefore, these mass values are not very distinctive predictions of the model. The more reliable and robust prediction is rather the values of the gauge couplings in the TeV-scale, given in Eq.~(\ref{coupling}), which do not change noticeably with the choice of the value of $M_R$ 
due to their logarithmic dependence on the energy scale.
  
Recall that our model is just the left-right model augmented by a colored scalar at the TeV-scale. Therefore, in similar to the usual left-right model it allows the right-handed neutrino $N_R$ to be Majorana in character. Although there is no mechanism that constraints right-handed neutrino mass $M_{N_R}$ in the left-right models, there exist bounds obtained from various low energy processes~\cite{Tello:2010am}. The LHC implications of TeV-scale left-right models regarding a heavy Majorana 
right-handed neutrino for variety of mass ranges have been studied in the literature~\cite{Mohapatra:2016twe,Helo:2013esa}. As for the future runs of the LHC; as recently studied in Ref.~\cite{Ruiz:2017nip}, for $g_R/g_L \sim 0.79$ (which is the case in our model as can be seen in Eq.~\eqref{coupling}), the 14 TeV LHC searches can probe the range $M_{W_R} \lesssim 6.3 - 7$ TeV for $M_{N_R}=100-700$ GeV.

\section{Phenomenology}
\label{sec:pheno}

\begin{figure}[!ht]
\centering
\includegraphics[scale=0.8]{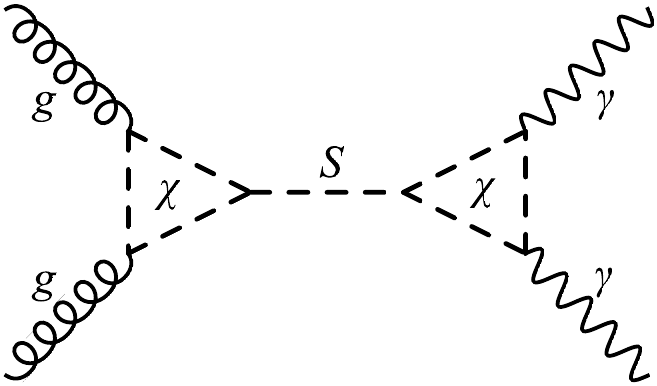}
\caption{The Feynman diagram of the production and decay of $\mc{S}$ at the LHC through $\chi$ in the loop.}
\label{fig:PD}
\end{figure}

In section~\ref{sec:mod}, we have discussed that the SM-singlet $\mathcal{S}$ can be as light as $\sim 1$ TeV and can potentially be observed at the LHC. 
Since $\mc{S}$ is a SM-singlet, it can not directly couple to the SM fermions and 
gauge bosons through any dimension-4 operator due to gauge invariance. 
Therefore, in order to produce $\mc{S}$ at the LHC, it is necessary to introduce 
extra colored particles that present in the loop. Similarly, for its decay to pair of EW 
gauge bosons, we need particles in the loop with nonzero hypercharge. These particles can be scalar, vector, or fermionic in nature. As mentioned previously, we would like to keep the matter and gauge sectors minimal and want to do a simplistic phenomenological study of that
scenario. We, therefore, choose only one colored and hypercharged scalar $\chi$ that appear naturally in our model and can serve
both the purposes, production and decay of $\mathcal{S}$ through loop interactions.
Note that the EM charge of $\chi$ is 4/3 which is the largest among the TeV-scale colored scalars in our model. Therefore, it couples to photon with a relatively greater strength which implies
large BR of $\mc{S}$ to diphoton. We further assume that $\chi$ is the lightest among all the colored and EM charged
scalars of our model and contributes most in our analysis. We neglect any small contamination from other particles in the 
loop assuming that they are heavier and thus, their effects are relatively suppressed.
In Fig.~\ref{fig:PD}, we present the Feynman diagram of the production of $\mc{S}$ from gluon-gluon fusion and 
its decay to two photons through $\chi$ in the loop.

\subsection{Production and decay}

\begin{figure}
\centering
\includegraphics[height=6cm,width=7.5cm]{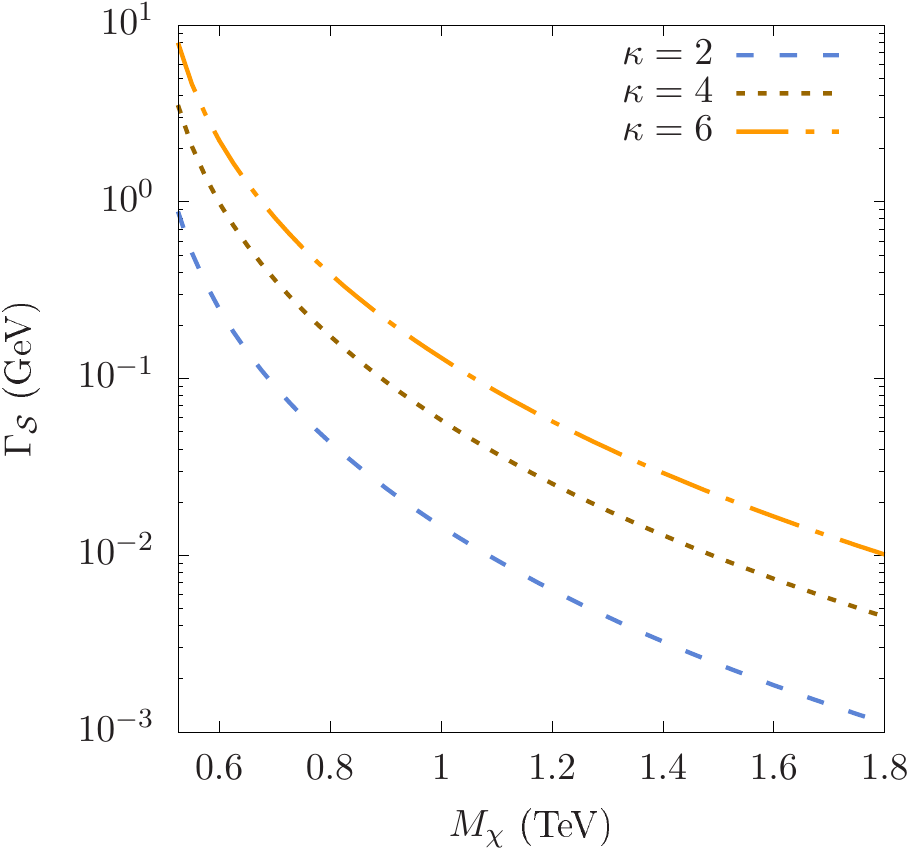}
\caption{Total width of $\mc{S}$ for $M_{\mc{S}}=1$ TeV as functions of $M_{\chi}$ for $\kp=2,4,6$ assuming $\Lm=5$ TeV.}
\label{fig:TW}
\end{figure}

The scalar $\mathcal{S}$ being singlet in nature, there is no
tree level couplings of $\mathcal{S}$ to the SM fermions and gauge bosons. It can decay to
a pair of SM gauge bosons only through nonrenormalizable dimension-5 operators. In the potential of the 
model, there could be some interaction terms which connect $\mathcal{S}$ with the SM Higgs doublet, which can lead to a mixing between $\mathcal{S}$ and the SM Higgs, after EWSB. Consequently, $\mathcal{S}$
can decay to a pair of the SM particles at the tree level. 
We know from experiments that the 125 GeV scalar observed at the LHC is very much the SM-like Higgs and therefore,
its mixing with $\mc{S}$ is expected to be small. For simplicity, we consider the $\mathcal{S}$-$h$ mixing, and therefore the partial widths of $\mathcal{S}$ 
to two SM fermions or two Higgs bosons, are negligible. Since $\chi$ carries color and hypercharge, it couples to the gluon and the $B_{\mu}$ (hypercharge) fields. Note that there is no coupling between $\chi$ and $W$ bosons, since $\chi$
is a singlet under $\mathrm{SU}(2)_L$. In the effective Lagrangian, we have the following dimension-5 operators
for the interactions of $\mathcal{S}$ with the SM gauge bosons prior to EWSB.
\begin{equation}
\label{eq:Lag}
\mathcal{L} \supset -\frac{1}{4}\kp_g~\mc{S}G_{\mu\nu}^aG^{a\mu\nu}
-\frac{1}{4}\kp_B~\mc{S}B_{\mu\nu}B^{\mu\nu}\ ,
\end{equation}
where $G_{\mu\nu}^a$ and $B_{\mu\nu}$ are the field-strength tensors for $\mathrm{SU}(3)_c$ and 
$\mathrm{U}(1)_Y$ gauge groups, respectively. Effective couplings $\kp_g$ and $\kp_B$ are associated with the gluon and the $B_{\mu}$ fields respectively. These couplings can be computed from the knowledge of the trilinear coupling related
with the $\mc{S}|\chi_i|^2$ interaction term. In general, for $N_f$ number of colored scalars $\chi_i$
with hypercharge $Y_i$ and for an interaction term $y_{\mc{S}}^i~\mc{S}|\chi_i|^2$, the
effective couplings are expressed as
\begin{align}
\kp_g &= \frac{\al_S}{2\pi}\lt|\sum_{i=1}^{N_f}\frac{1}{6}C_R^i\frac{y_{\mc{S}}^i}{M_{\chi_i}^2}I_0\lt(\frac{4M_{\chi_i}^2}{M_{\mc{S}}^2}\rt)\rt|\ , \\
\kp_B &= \frac{\al}{2\pi c_W^2}\lt|\sum_{i=1}^{N_f}\frac{1}{6}d_R^i\lt(\frac{Y_i}{2}\rt)^2\frac{y_{\mc{S}}^i}{M_{\chi_i}^2}I_0\lt(\frac{4M_{\chi_i}^2}{M_{\mc{S}}^2}\rt)\rt|\ ,
\end{align}
where $d_R^i$ is the dimension of the $\mathrm{SU}(3)$ representation ({\it e.g.} $d_R=3$ for triplet and $d_R=8$ for octet representations)
and $C_R^i$ is the index of the $\mathrm{SU}(3)$ representation ({\it e.g.} $C_R=1/2$ for triplet and $C_R=3$ for octet representations). The strong and the electromagnetic couplings are denoted by $\al_S$ and $\al$, respectively. The cosine of
the Weinberg angle is denoted as $c_W$. The loop function $I_0$ is given by,
\begin{equation}
I_0(\tau) = -3\tau\lt[1-\tau\lt\{\sin^{-1}\lt(\frac{1}{\sqrt{\tau}}\rt)\rt\}^2\rt]\ .
\end{equation}
For only one colored triplet and hypercharged ($Y=8/3$) scalar $\chi$, $N_f=1$, $d_R=3$ and
$C_R=1/2$. To keep our results as model independent as possible, we assume $y_{\mc{S}} = \kp \Lm$, where $\Lm$ is some new physics scale (this can be chosen as $M_R$) for which
we choose 5 TeV for all our computations and we keep $\kp$ as a free parameter. The BRs of 
$\mc{S}$ to $gg$, $\gm\gm$, $Z\gm$ and $ZZ$ modes are 90.6\%, 5.6\%, 3.3\% and 0.5\% respectively.

Here, we assume that $M_{\chi} > M_{\mc{S}}/2$, and therefore $\mc{S}$ cannot decay to a $\chi$ pair.
It is important to note that the BR depends only on $Y$, not on the other parameters. This is because
all the partial widths, and hence the total width, scale as $\kp^2\Lm^2$ and the loop function $I_0$ 
(for any values of $M_{\mc{S}}$ and $M_{\chi}$) would be the same for all the partial widths. The scalar $\mc{S}$ has the largest BR in the dijet channel. We expect the BR in the $\gm\gm$, $Z\gm$ and $ZZ$ are of
similar order, but $ZZ$ mode is suppressed due to its phase space factor. The total width 
$\Gm_{\mc{S}}$ is a function of $M_{\mc{S}}$, $M_{\chi}$ and $\kp\Lm$. In Fig.~\ref{fig:TW}, we
show $\Gm_{\mc{S}}$ as functions of $M_{\chi}$ for $M_{\mc{S}}=1$ TeV for three different values of 
$\kp$ assuming $\Lm=5$ TeV. As mentioned previously, $\Gm_{\mc{S}}$ scales as $\kp^2\Lm^2$ and one can
easily estimate the total width for other values of $\kp\Lm$ from this plot.

\subsection{Exclusions from LHC data}

To derive bounds on the model parameters from the LHC data and related numerical analysis, we implement the Lagrangian given in Eq.~(\ref{eq:Lag})
in {\sc FeynRules2.0}~\cite{Alloul:2013bka} to generate the model files for the
{\sc MadGraph5}~\cite{Alwall:2014hca} event
generator. We use CTEQ6L1~\cite{Pumplin:2002vw} parton distribution functions (PDF)
to compute cross sections. We fix the factorization and renormalization scales
at $M_{\mc{S}}$ for all our numerical computations.

\begin{figure*}
\centering
\subfloat[]{\includegraphics[height=6.1cm,width=6.5cm]{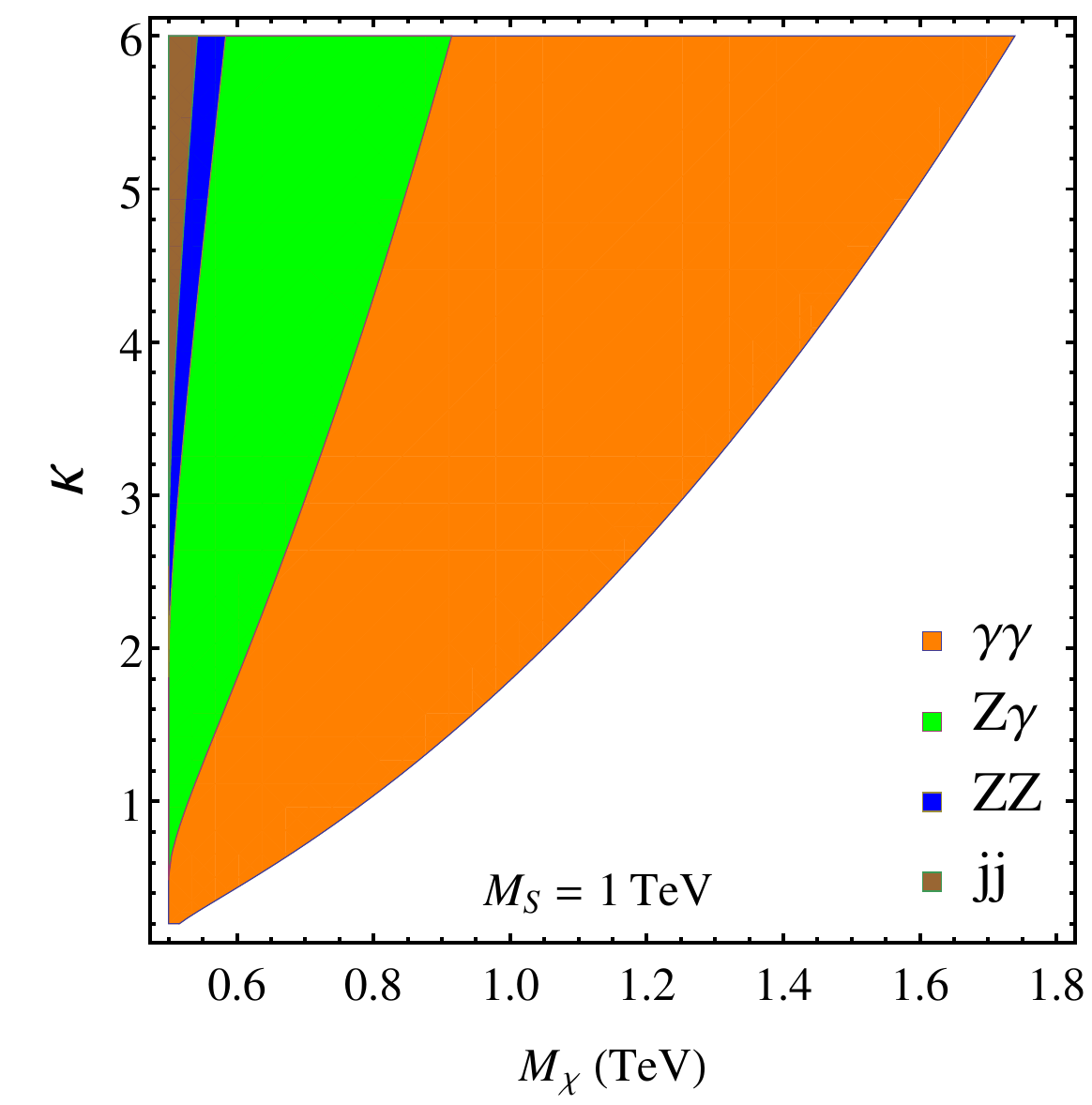}\label{fig:kMchi}}~~~~
\subfloat[]{\includegraphics[height=6cm,width=6.5cm]{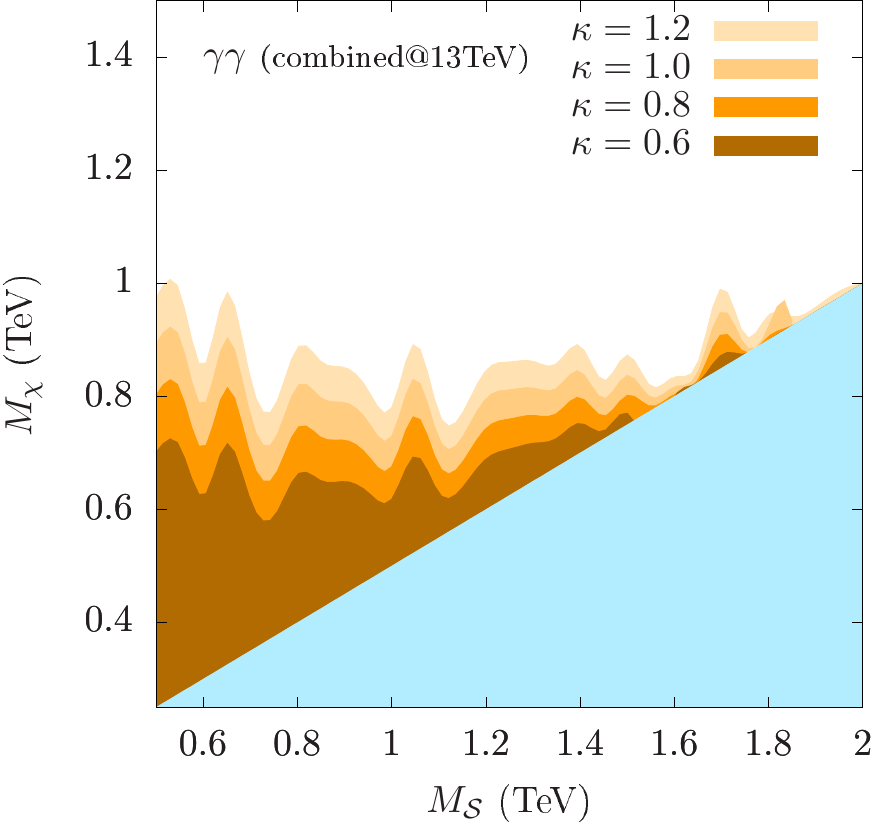}\label{fig:comb}}
\caption{(a) The excluded region in $M_{\chi}-\kp$ plane for $M_{\mc{S}}=1$ TeV from the 13 TeV LHC data. The orange, green, blue and brown regions represent the region ruled out by the $\gm\gm$, 
$Z\gm$, $ZZ$ and $jj$ resonance search data respectively. The excluded region in $M_{\mc{S}}-M_{\chi}$ plane for different $\kp$ with $\Lm=5$ TeV
using the 13 TeV combined ATLAS and CMS diphoton resonance search. ATLAS and CMS data are combined
statistically using Eq.~\eqref{eq:comb}. The widths of the resonance assumed by ATLAS and CMS are 4 MeV~\cite{ATLAS:2016eeo} and 140 MeV~\cite{Khachatryan:2016yec}. The sky-blue region cannot be probed in our analysis as we assume $M_{\chi} > M_{\mc{S}}/2$.}
\label{fig:bound}
\end{figure*}

For our phenomenological analysis, we have only three free parameters {\it viz.}
$M_{\mc{S}}$, $M_{\chi}$ and $\kappa$ (we choose $\Lm=5$ TeV for all our numerical computations).
We first derive bounds on the parameters from the latest LHC 13 TeV $\gm\gm$~\cite{ATLAS:2016eeo,Khachatryan:2016yec}, 
$Z\gm$~\cite{CMS:2016pax,CMS:2016cbb}, $ZZ$~\cite{ATLAS:2016npe} and 
$jj$~\cite{ATLAS:2016lvi,CMS:2016wpz} resonance search data.
The observed upper limit (UL) at 95\% confidence level (CL) on the cross sections for the resonance mass of 1 TeV of four type of resonances are given by,
\begin{align}
\sg_{\gm\gm} &\lesssim 1~\textrm{fb},~\sg_{Z\gm} \lesssim 10~\textrm{fb},\nn\\
\sg_{ZZ} &\lesssim 20~\textrm{fb},~\sg_{jj} 
\lesssim 7.5~\textrm{pb}.
\end{align}  
These values are used in Fig.~\ref{fig:kMchi} where we show the excluded parameter space (colored regions) in 
$M_{\chi}-\kp$ plane for $M_{\mc{S}}=1$ TeV. The excluded regions shown in orange, green, blue and
brown are derived from the $\gm\gm$, $Z\gm$, $ZZ$ and $jj$ resonance search data. We can see that the diphoton data is the most powerful in 
constraining the parameter space in $M_{\chi}-\kp$ plane. In Fig.~\ref{fig:comb}, we
present the excluded regions in $M_{\mc{S}}-M_{\chi}$ plane for different $\kp$ with 
$\Lm=5$ TeV from the latest 13 TeV combined ATLAS and CMS diphoton resonance search data.
Cross section ULs ($\sg_i$) from different experiments and the corresponding uncertainties ($\Dl\sg_i$) are combined statistically using the following relations,
\begin{align}
\label{eq:comb}
\frac{1}{(\Dl\sg_c)^2}=\displaystyle\sum_i\frac{1}{(\Dl\sg_i)^2};~~
\frac{\sg_c}{(\Dl\sg_c)^2}=\displaystyle\sum_i\frac{\sg_i}{(\Dl\sg_i)^2}\ ,
\end{align}
where $\sg_c$ is the combined cross section and $\Dl\sg_c$ is the 
uncertainty associated with it. In case of asymmetric uncertainties, we get
$\Dl\sg_i$ by averaging upper and lower uncertainties. Although uncertainties are used to
compute $\sg_c$, we have not shown the uncertainty bands in the exclusion plots for simplicity.
The sky-blue regions in these plots cannot be probed in our set-up as we
always assume $M_{\chi} > M_{\mc{S}}/2$. If $M_{\chi} < M_{\mc{S}}/2$, the $\mc{S}\to \chi\chi$
decay becomes kinematically allowed and becomes the dominant decay
mode of $\mc{S}$. This will make the diphoton and other branching modes suppressed. Therefore, observing 
$\mc{S}$ in the $\gm\gm$, $Z\gm$, $ZZ$ and $jj$ resonance searches become much more challenging.
One should note that exclusion regions are not very sensitive to the $M_{\chi}$ values for a fixed 
$\kp$. This is because for heavier resonances, the cross section ULs are not very sensitive to the 
resonance mass due to lack of statistics and therefore, the quantity $\sg(M_{\mc{S}},M_{\chi})\times BR$ 
should remain insensitive for heavier resonances. The reduction in the production cross section as we increase $M_{\mc{S}}$ 
is compensated by the slight change in $M_{\chi}$ since $\sg(M_{\mc{S}},M_{\chi})$ quantity is very sensitive
to the $M_{\chi}$.
Note that these bounds are derived from the 
observed 95\% CL UL on the cross sections. Consideration of uncertainties on the 
cross sections limits would relax the derived bound somewhat. 
In all our computations, we have considered a next-to-leading order $K$-factor of 2 
to account for the higher-order effects~\cite{deFlorian:2016spz}.

\subsection{Future prospects at the LHC}

\begin{figure*}
\centering
\includegraphics[height=9cm,width=16cm]{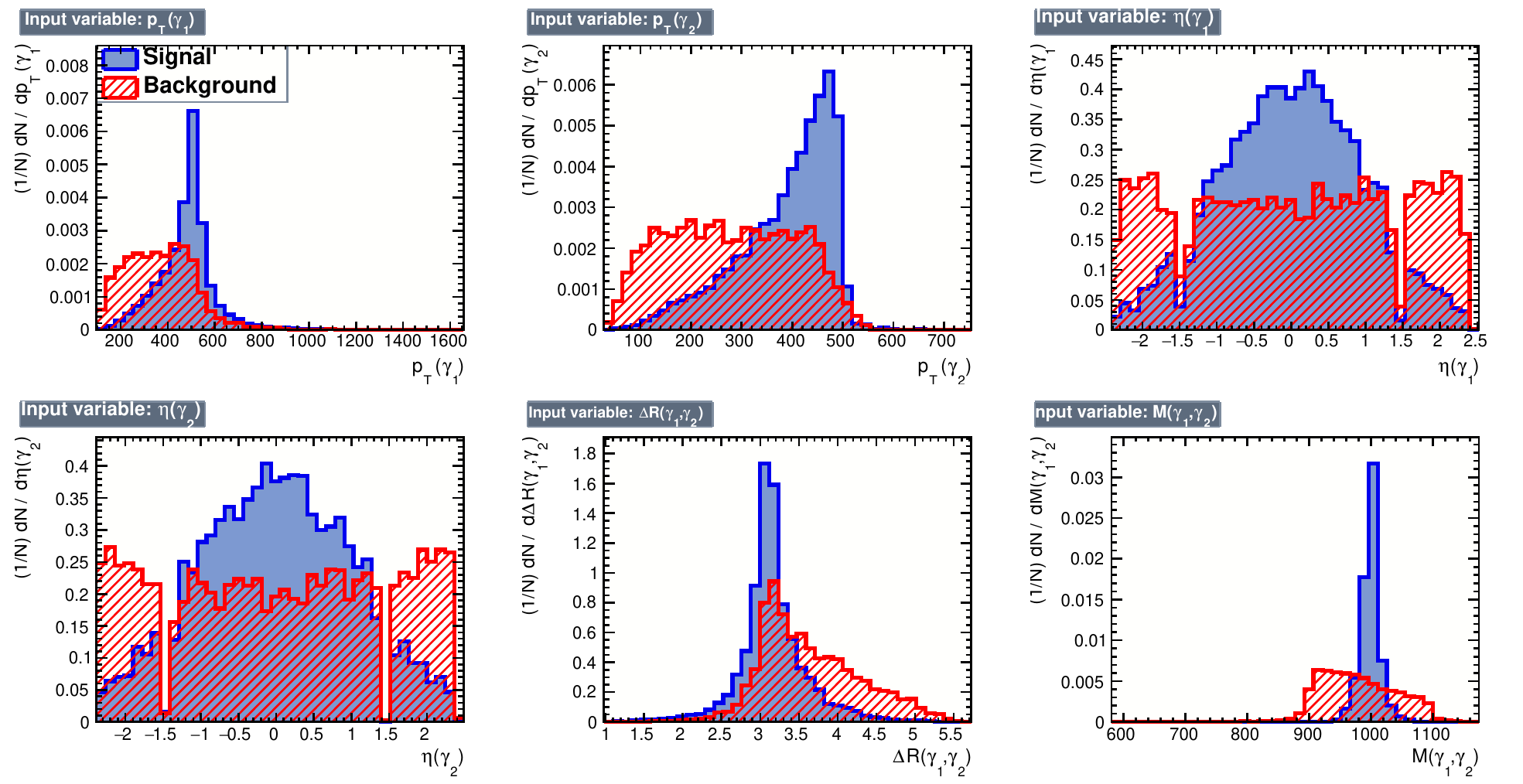}
\caption{Signal (blue) and background (red) distributions of the input variables used for MVA. These
distributions are generated for $M_{\mc{S}}=1$ TeV assuming $\Gm_{\mc{S}}=1$ GeV. We do not present the jet multiplicity distribution
here since it has the smallest RI in the MVA as shown in \ref{tab:RI} and therefore, the $N_{jet}$ distribution would not differ much for the signal and the background. }
\label{fig:BDTvar}
\end{figure*}

\begin{table*}
\setlength{\tabcolsep}{12pt}
\centering
\begin{tabular}{|c|c|c|c|c|c|c|c|}
\hline 
Variables & $p_T(\gm_1)$ & $p_T(\gm_2)$ & $\eta(\gm_1)$ & $\eta(\gm_1)$ & $\Dl R(\gm_1,\gm_2)$ & $M(\gm_1,\gm_2)$ & $N_{jet}$ \\ 
\hline 
$\mathrm{RI}\times 10^{-1}$ & 1.22 & 1.31 & 1.02 & 1.11 & 1.30 & 3.25 & 0.78 \\ 
\hline 
\end{tabular} 
\caption{\label{tab:RI} Input variables used for MVA to separate the signal from the background and their relative importances (RIs). These numbers are shown for $M_{\mc{S}}=1$ TeV.}
\end{table*}

\begin{figure*}
\centering
\subfloat[]{\includegraphics[height=5.5cm,width=6.5cm]{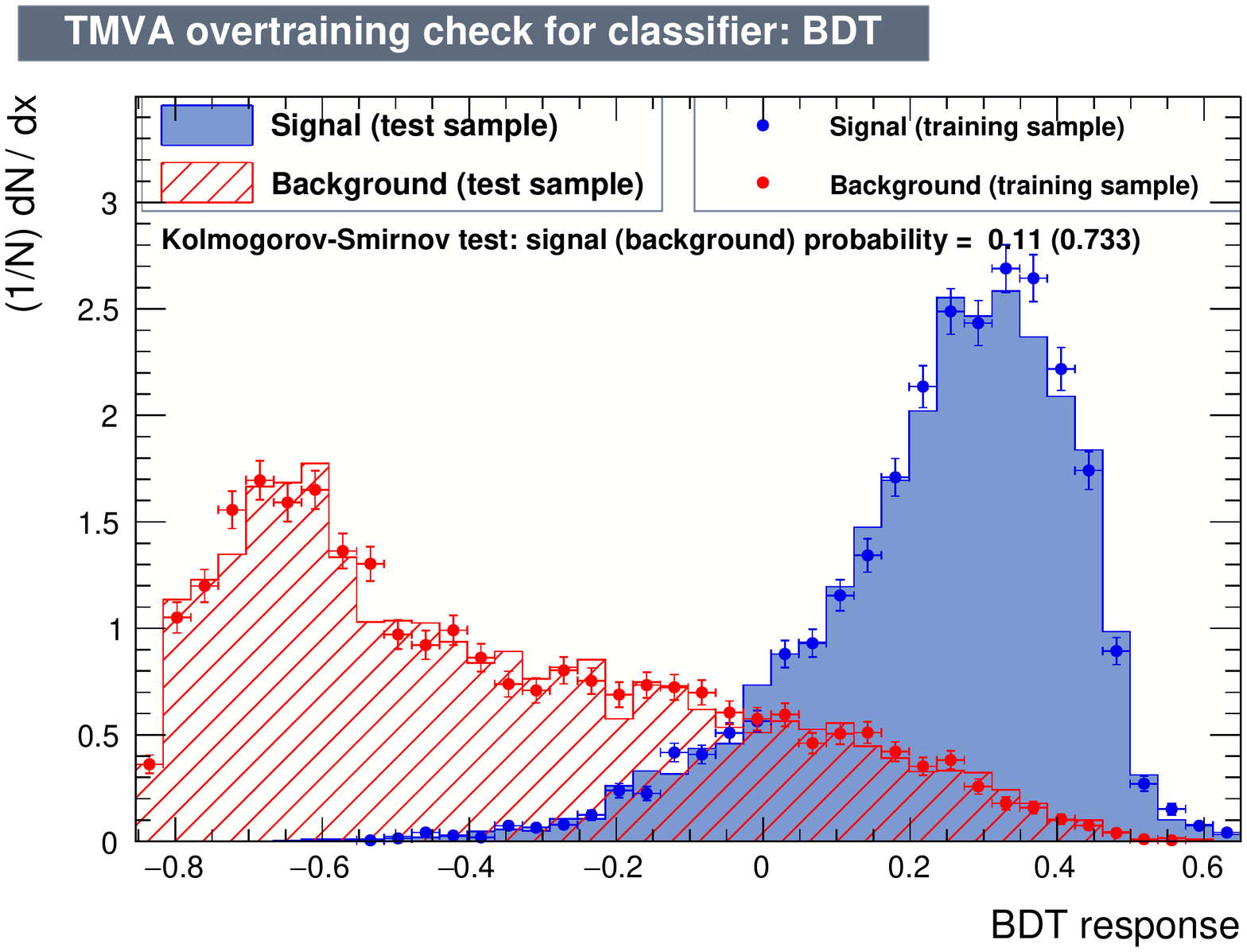}\label{fig:BDTres}}~~~
\subfloat[]{\includegraphics[height=5.5cm,width=6.5cm]{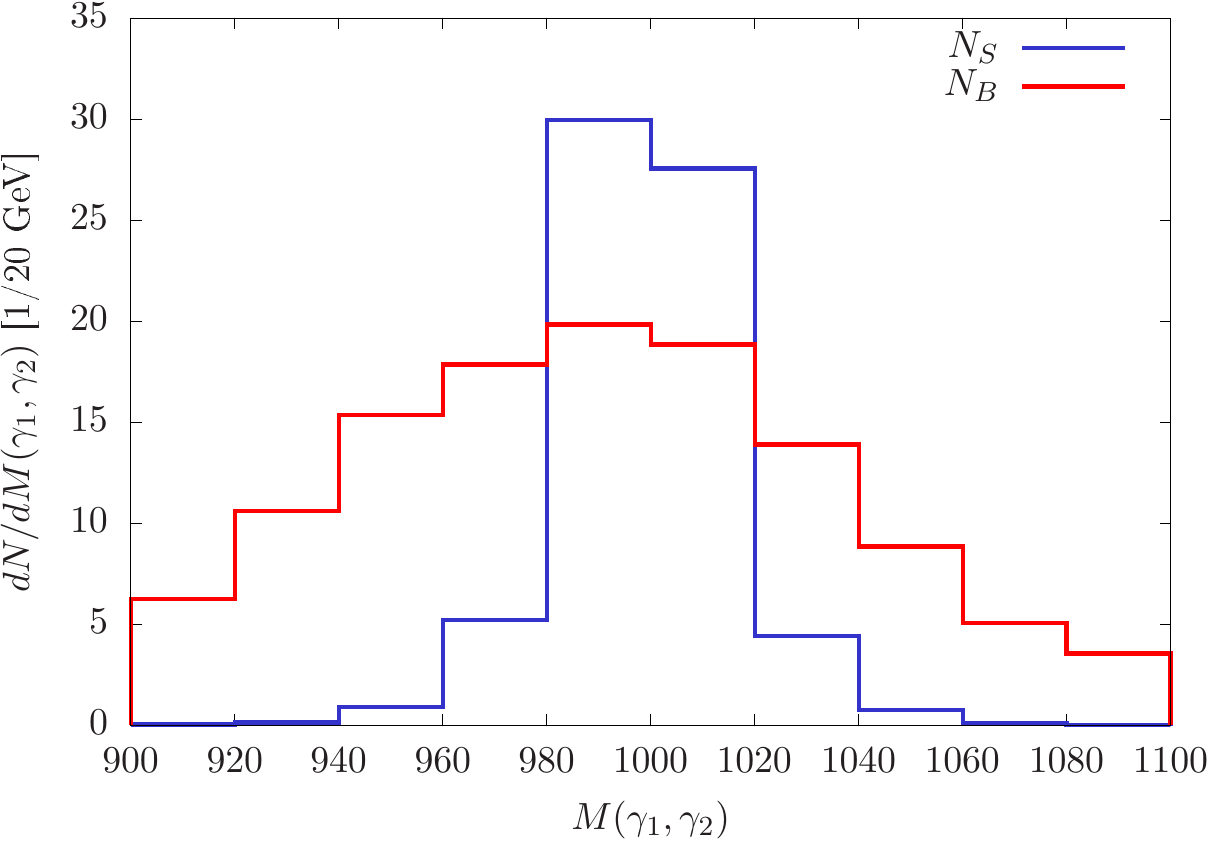}\label{fig:Myy}}
\caption{BDT response of the signal (blue) and the background (red) for $M_{\mc{S}}=1$ TeV.
(b) $M(\gm_1,\gm_2)$ distributions for the signal (blue) and the background (red) for $M_{\mc{S}}=1$ TeV after 
applying the optimal BDT cut at $\sim 0$ to obtain $5\sg$ significance at the 13 TeV LHC with $\mc{L}=300$ fb$^{-1}$.}
\label{fig:BDT}
\end{figure*}

\begin{figure*}
\centering
\subfloat[]{\includegraphics[height=5.5cm,width=6.5cm]{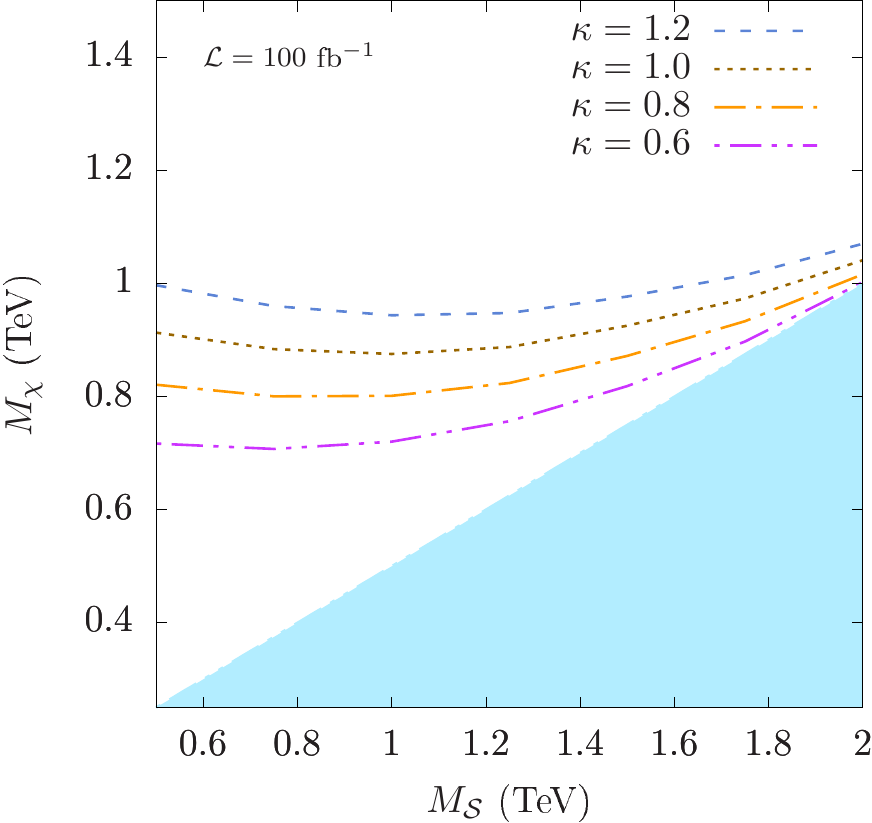}\label{fig:pros100}}~~~
\subfloat[]{\includegraphics[height=5.5cm,width=6.5cm]{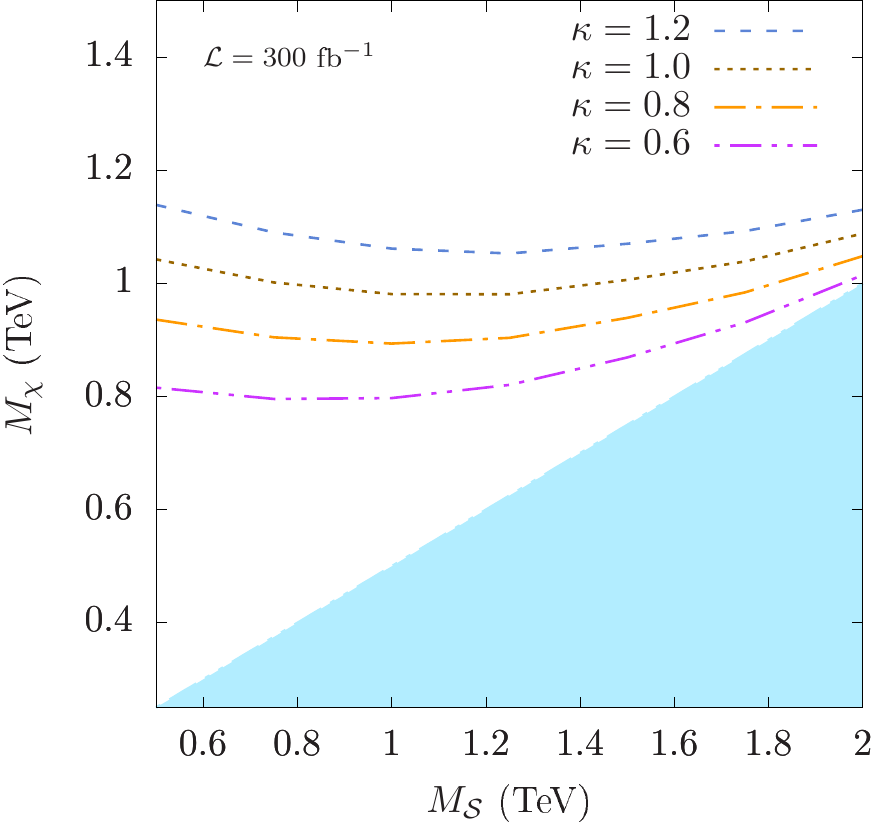}\label{fig:pros300}}\\
\subfloat[]{\includegraphics[height=5.5cm,width=6.5cm]{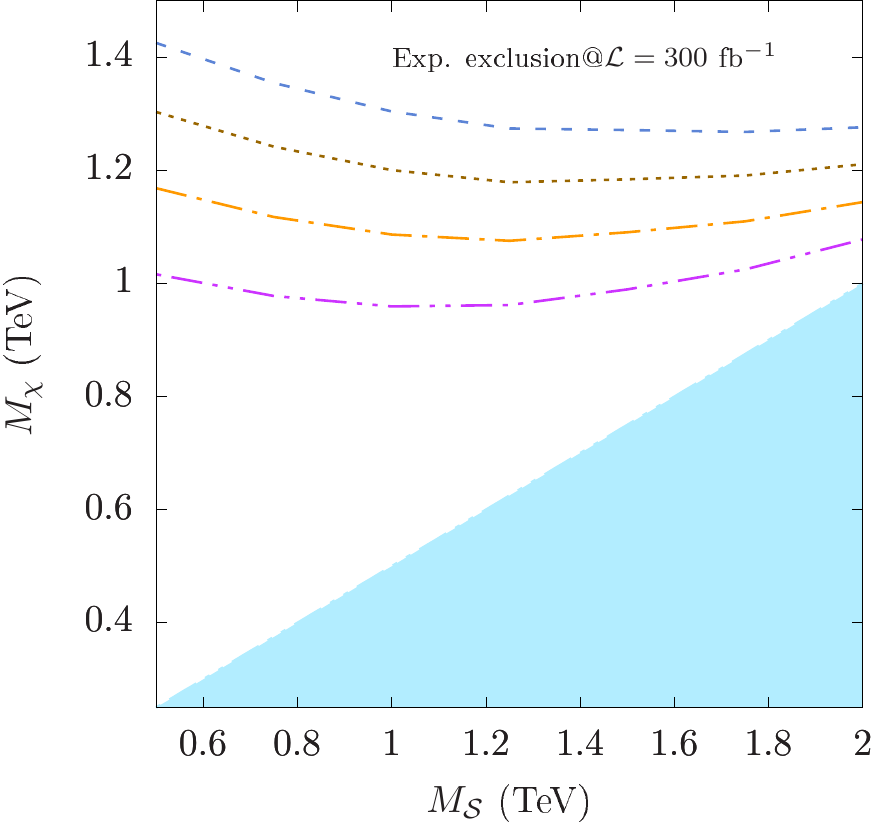}\label{fig:sen300}}
\caption{The $5\sg$ discovery contours in $M_{\mc{S}}-M_{\chi}$ plane for different $\kp$
at 13 TeV LHC for (a) 100 fb$^{-1}$ and (b) 300 fb$^{-1}$ integrated luminosity. (c) The expected $2\sigma$ exclusion plot for $\mathcal{L}=300$ fb$^{-1}$. The sky-blue region cannot be probed in our analysis as we assume $M_{\chi} > M_{\mc{S}}/2$.}
\label{fig:pros}
\end{figure*}

In this subsection, we look at the prospect to discover $\mc{S}$ at the 13 TeV 
LHC runs with high integrated luminosities. In previous subsection, we find that 
the most stringent bounds come from the diphoton data. Therefore, we only focus 
on the diphoton final state for the present prospect study. After event generation, 
we use \textsc{Pythia6}~\cite{Sjostrand:2006za} for parton shower and hadronization. 
The subsequent detector simulation is done using \textsc{Delphes3}~\cite{deFavereau:2013fsa}
package. Jets are clustered with \textsc{FastJet}~\cite{Cacciari:2011ma} using
the anti-$k_T$ algorithm~\cite{Cacciari:2008gp} with the clustering parameter, 
$R = 0.4$. We use \textsc{TMVA}~\cite{Hocker:2007ht} for the multivariate analysis.

Signal events are generated with up to two jets {\it i.e.} $pp\to \mc{S}(\to\gm\gm)+0,1,2$~jets
which are MLM~\cite{Mangano:2006rw} merged at a matching scale $Q_{cut}=50$ GeV.
The dominant (roughly 90\%) SM background for this signal comes from the $qq\to\gm\gm$ process.
Similar to the signal, we generate this background by merging $pp\to\gm\gm+0,1,2$~jets processes
at $Q_{cut}=15$ GeV. We only consider this dominant background in our analysis. 
Appropriate matching scales for signal and background are determined by assuring
smooth transition in the differential jet-rate distributions between events with $N$ and $N + 1$ jets
and matched cross sections are within $\sim 10\%$ of the zero jet contribution. We also
check the stability of the matched cross section with the variation of $Q_{cut}$ once
it is properly chosen.

The 13 TeV diphoton data already set an UL on $\sg\times BR\sim 1$ fb for the 
resonance mass of around 1 TeV. Therefore, it is very challenging to observe 
such a signal over the large SM background. ATLAS and CMS collaborations use
cut-based technique in their diphoton resonance searches at the 13 TeV LHC. 
In this paper, to obtain better sensitivity, we use a MVA to discriminate tiny signal from the large SM background. ATLAS and CMS ULs on $\sg\times BR$ slightly depend on the width of the resonance
but we use a fixed width of 1 GeV for all $M_{\mc{S}}$ in the following analysis for simplicity.
The width of $\mc{S}$ is a function of model parameters viz. $M_{\mathcal{S}}$, $M_{\chi}$ and $\kappa\Lambda$. Instead of choosing
a specific benchmark, we use $\Gamma_{\mathcal{S}}=1$ GeV for our MVA. This analysis is insensitive 
to the actual width choice as long as $\Gamma_{\mathcal{S}}\ll M_{\mathcal{S}}$ i.e. the narrow width approximation is well-valid.

We generate signal and background events with some basic transverse momentum 
($p_T$), pseudorapidity ($\eta$) and separation in $\eta-\phi$ plane ($\Dl R$) 
cuts as follows: 
\begin{align}
p_T(x)>25~\mbox{GeV},~|\eta(x)|<2.5,~\Dl R(x,y) >0.4
\end{align}
where $x,y=\{\gm,j\}$. We use a strong selection cut on the invariant mass 
of the photon pair, $|M(\gm\gm)-M_{\mc{S}}|<100$ GeV to reduce the huge diphoton 
background before passing events to TMVA. For MVA, we use the Boosted Decision 
Tree (BDT) algorithm where we feed the following seven kinematic variables: 
$p_T(\gm_1)$, $p_T(\gm_2)$, $|\eta(\gm_1)|$, $|\eta(\gm_2)|$, $\Dl R(\gm_1,\gm_2)$, 
$M(\gm_1,\gm_2)$ and jet multiplicity ($\gm_1$ and $\gm_2$ are the two selected photons
ordered according to their $p_T$). In Fig.~\ref{fig:BDTvar}, we show the 
signal (blue) and background (red) distributions of these variables used in MVA. 
We choose these simple variables which
are less correlated and have sufficiently good discriminating power. In Table
\ref{tab:RI}, we show the relative importance (RI) of these variables for the
benchmark mass $M_{\mc{S}}=1$ TeV. We find that the two variables $M(\gm_1,\gm_2)$ 
and $\Dl R(\gm_1,\gm_2)$ are very effective in discriminating signal from background. 
Other variables like $p_T$ and $\eta$ of photons also have reasonably good discriminating power.
We obtain the cut efficiency of almost 75\% for the signal but as small as 10\%
for the background for the whole range of $M_{\mc{S}}$ we considered. It is important 
to mention that this set of seven variables used might not be the optimal one. There 
is always a scope to improve the analysis with cleverer choices of variables.

The BDT algorithm is prone to overtraining and therefore, one should always
be careful while using it in MVA. Overtraining of the signal and background test samples
can usually happen due to the improper choices of BDT tuning parameters. 
Whether a test sample is overtrained or not can be checked by using the 
Kolmogorov-Smirnov (KS) statistics. Generally, a test sample is not overtrained 
if the corresponding KS probability lies within the range 0.1 to 0.9. In our
analysis, we use two statistically independent samples for each $M_{\mc{S}}$ choice,
one for training and the other for testing the BDT. In Fig.~\ref{fig:BDTres}, we 
show the BDT response of the signal and background for the benchmark mass
$M_{\mc{S}}=1$ TeV. From the BDT response, one can see that a BDT cut around $\sim 0$ 
can effectively separate the signal from the background and lead to best significance.
In Fig.~\ref{fig:Myy}, we show the $M(\gm_1,\gm_2)$ distributions for the signal and the
background for $M_{\mc{S}}=1$ TeV at the 13 TeV LHC with $\mc{L}=300$ fb$^{-1}$.
This plot is shown for the significance of $5\sg$ where $\sg=N_S/\sqrt{N_S+N_B}$ and the number of 
signal and background events that survive after the optimal BDT cut ($>0$) are $N_S=69$ with cut efficiency 0.75 and $N_B=120$ with cut efficiency 0.1 respectively.

In Figs.~\ref{fig:pros100} and ~\ref{fig:pros300}, we show the $5\sg$ discovery contours in $M_{\mc{S}}-M_{\chi}$ plane 
for different $\kp$ at the 13 TeV LHC for 100 and 300 fb$^{-1}$ integrated luminosities respectively. 
As stated earlier, the sky-blue region {\it i.e.} $M_{\chi} < M_{\mc{S}}/2$ 
is not considered in our analysis. We observe that the discovery reach for 100 fb$^{-1}$
run in Fig.~\ref{fig:pros100} is not much improved from the bounds obtained in Fig.~\ref{fig:bound}.
But for 300 fb$^{-1}$ run, a substantially bigger region of parameter space can be probed.
In Fig.~\ref{fig:sen300}, we show the expected 95\% CL exclusion plot in $M_{\mc{S}}-M_{\chi}$ plane 
for $\mathcal{L}=300$ fb$^{-1}$. It is obvious that the parameter space which can be excluded with
95\% CL is much bigger than the parameter space which can be discovered with $5\sigma$ significance.
As previously mentioned, the limits on $\sg\times BR$ for a scalar decays to diphoton are already
very strict. Therefore, to observe such a scalar at the LHC is very challenging and we need a more dedicated analysis for that.

In this paper, we choose to use a MVA for the LHC prospect study to achieve better sensitivity 
to the parameter space compared to a cut-based analysis. To give the readers a rough idea of gain in sensitivity, we wish to present here a quantitative comparison 
between the two types of analyses for the benchmark mass $M_{\mc{S}}=1$ TeV. 
We apply further the following hard cuts on photons viz. $p_T(\gamma_1),p_T(\gamma_2)> 200$ GeV
and $|M(\gamma_1,\gamma_2) - M_{\mathcal{S}}|<50$ GeV on the events that are used for the BDT analysis.
In context of Fig.~\ref{fig:Myy}, we have discussed previously that the number of signal and background 
events which survive after the optimal BDT cut (around $\sim 0$) are 69 and 120 respectively. The corresponding 
signal and background events that survive after the cut-based analysis are 65 and 432 respectively which
leads to a $\sim 3\sg$ significance. One can see, therefore, that BDT analysis is very effective in terms of background reduction
compared to a cut-based analysis. Note that this set of cuts is not fully optimized (but fairly good)
and one can vary these cuts to find the optimized set of cuts to improve the significance from 
$\sim 3\sigma$. But an optimized BDT analysis is always expected to perform better than an optimized
cut-based analysis as long as a clever set of variables are used.
A BDT analysis is usually more effective than a cut-based analysis especially in the low mass (here low $M_{\mc{S}}$) region.
For heavier masses, where the SM background is expected to be very small compared to the signal, an optimized cut-based analysis can compete to an optimized BDT analysis. 


\section{Summary}
\label{summary}
In this paper, we explore the phenomenology of TeV-scale scalars in the non-supersymmetric 
$\mathrm{SO}(10)$ grand unification framework. In particular, we investigate the LHC phenomenology
of a SM singlet scalar $\mc{S}$ which interacts with gluons and photons through loop interactions
with a color-triplet hypercharged scalar $\chi$ which is remnant from the breaking of the Pati-Salam gauge group, $\mathrm{SU}(2)_L\otimes \mathrm{SU}(2)_R\otimes \mathrm{SU}(4)_C$. The part of the model that lies in the TeV-scale is the left-right model, whose gauge group is $\mathrm{SU}(2)_L\otimes \mathrm{SU}(2)_R\otimes \mathrm{U}(1)_{B-L}\otimes \mathrm{SU}(3)_C$, augmented with the color-triplet scalar  $\chi$. The scalar 
$\mc{S}$ is a component of an $\mathrm{SU}(2)_R$ triplet scalar which is responsible for the breaking of the left-right model into the SM. Note that we have stayed in the minimal picture in terms of the total field content; the model does not have any extra matter fields or any $\mathrm{SO}(10)$ multiplets in the scalar content other than the ones required to begin with. 

The colored scalar in our set-up effectively induces the interaction terms of $\mc{S}$ with gluons and photons that lead to a diphoton final state after being produced via gluon fusion. In addition to the $\gm\gm$ decay, $\mc{S}$ can also decay to $jj$,
$\gm Z$ and $ZZ$ modes. We present the exclusion region in $M_{\chi}-\kp$ plane for a benchmark resonance mass $M_{\mc{S}}=1$ TeV using the latest LHC data. We find that the most stringent bounds on the parameter space of our model come from the diphoton resonance search data. Therefore, we consider the diphoton channel as the most promising channel for the discovery of $\mc{S}$
at the LHC. As a prospect study, we compute the higher-luminosity LHC discovery reach of $\mathcal{S}$ by using a
state-of-the-art multivariate technique. We present $5\sg$ discovery contours for different $\kp$ choices in the 
$M_{\mc{S}}-M_{\chi}$ plane at the 13 TeV LHC with 100 and 300 fb$^{-1}$ integrated luminosity. From our analysis, we find that for $\kp\sim 1$, $M_{\mc{S}}\sim 0.5-2$ TeV and $M_{\chi}\sim 1$ TeV can easily be observed with $5\sg$ confidence level at the 13 TeV LHC with 300 fb$^{-1}$ integrated luminosity. 
Note that the role of various systematic uncertainties 
are always important to consider in an analysis for robust and accurate prediction. But in the current scope, we do not consider systematic uncertainties for simplicity. 
 
The unifications of the couplings in the model are successfully realized, where the TeV-scale colored-triplet plays an important role. As discussed in~\cite{Aydemir:2015oob}, it is very difficult to achieve a successful $\mathrm{SO}(10)$ grand unification set-up with a TeV-scale left-right model. Slightly modifying the low energy scalar content by allowing a number of colored scalars, originated from various Pati-Salam multiplets, to become light generates the possibility to accommodate a TeV-scale left-right model in the $\mathrm{SO}(10)$ grand unification framework. Among a number of low energy scalar configurations, the ones with the very color-triplet selected in our model appear to particularly stand out~\cite{Aydemir:2015oob}. We also note that the values obtained for the intermediate scale (where the Pati-Salam is broken) and the unification scale are sufficiently high to remain compatible with the experimental constraints regarding the leptoquark induced effects and the proton decay. 

\section*{Acknowledgements} 

We would like to thank Ilia Gogoladze for stimulating conversations. This work is supported by the Swedish Research Council under contract 621-2011-5107. T.M. is supported by the Carl Trygger Foundation under contract CTS-14:206.

\bibliography{reference}{}
\bibliographystyle{unsrt}

\end{document}